\documentclass[aps,prb,twocolumn,superscriptaddress,showpacs,floatfix]{revtex4}
\usepackage{graphicx}
\usepackage{latexsym}
\usepackage{stmaryrd}
\usepackage{amsmath}
\usepackage{epstopdf}

\begin{document}
\title{Numerical study of Klein quantum dots in graphene system}
\author{Jiaojiao Zhou}
\affiliation{College of Physics, Optoelectronics and Energy, Soochow University, Suzhou, 215006, China}
\author{Shu-guang Cheng}
\affiliation{Department of Physics, Northwest University, Xi'an 710069, People's Republic of China}
\author{Hua Jiang}\email{jianghuaphy@suda.edu.cn}
\affiliation{College of Physics, Optoelectronics and Energy, Soochow University, Suzhou, 215006, China}
\affiliation{Institute for Advanced Study, Soochow University, Suzhou, 215006, China}

\date{\today}

\begin{abstract}
Klein quantum dot (KQD) refers to a QD with quasi-bound states and a finite trapping time, which has been observed in experiments focused on graphene recently. In this paper, we develop a numerical method to calculate local density of states (LDOS) of KQD and apply it to monolayer graphene. By investigating the variation of LDOS in a circular quantum dot, we obtain the dependence of the quasi-bound states on the quantum dot parameters (e.g. the electron energy, radius, confined potential, etc). Based on these results, not only can we well explain the experimental phenomena, but also demonstrate how quasi-bound states turn to real bound states when intervalley scattering is taken into considered. We further study the evolution of the LDOS for KQD varying from a circle shape to a semicircle shape, which reveals the mechanism of whispering gallery mode on the quasi-bound states.
\end{abstract}


\maketitle

\section{Introduction}
The study of zero dimensional (0D) bound states has become an important topic as quantum dot. By decreasing the sample's size or applying a localized potential field in a semiconductor system\cite{BaconM,zhouW,ZhongH,LeiY,LeonardD,PengX,DeFS}, 0D bound states can be achieved by confining the carriers in a nanoscale region. However, in the graphene systems, carriers cannot be completely confined by potential because of the Klein tunneling \cite{CalogeracosA,CheianovVV, KatsnelsonMI}. Therefore, the early experiments tend to focus on the fabrication of small samples in graphene sheet to obtain 0D bound states \cite{HamalainenSK, PharkSH, PonomarenkoLA, StampferC, SubramaniamD}. These experiments could be highly expensive and technically challenging, which may limit its practical application.

Recently, according to the new recognition of the bound states in the graphene systems \cite{SilvestrovPG,ChenHY}, scientists find that the quasi-bound states can be obtained by applying a local potential, which is quite different from the previous understanding of bound states. The region where the quasi-bound states localized is called KQD \cite{GutierrezC, JolieW, LeeJ}. These new quasi-bound states can be used to fabricate graphene QD, which is technically much easier than decreasing the size of graphene samples. On the boundary of graphene KQD, the oblique incident massless electrons will be reflected with high probability. Moreover, for some special energies, the reflected electrons can even construct interference with themselves after multiple reflections and therefore form the quasi-bound states.  During the whole processes, the massless electrons behave like acoustic waves. Such mechanism is similar to whispering gallery mode, where the incident and reflected wave, with fixed frequency, interfere with each other and form standing wave inside the circular cavity\cite{GhahariF,ZhaoY,ForemanMR}. In the last two years, many experimental groups  demonstrated the existence of KQD in graphene systems, such as the heterostructures of graphene/hexagonal boron nitride\cite{GhahariF,ZhaoY,LeeJ,VelascoJ,FreitagNM}, graphene/metal Cu(111)\cite{GutierrezC,BaiKK2,BaiKK} and graphene/metal Mo\cite{QiaoJB}. These novel observations have attracted lots of attentions.

The previous theoretical study of KQD is  based on solving Dirac equation, where LDOS is obtained to characterize the quasi-bound state\cite{BardarsonJH, CsertiJ, DowningCA, MatulisA, SchulzQ,SchulzP,WuJS}. The properties of circular KQD can be obtained by adopting this method, and the theoretical observation is basically consistent with the experiments results\cite{GutierrezC,LeeJ,BaiKK}. However, the method still has some limitations. Firstly, it only considers a single-valley structure and ignores the effect of intervalley scattering. According to our previous study, the intervalley scattering exists in the step-changed potential interface\cite{zhouJ}. Secondly, many detailed factors which experimentally do exist cannot be included in this method, such as strain field\cite{ZhuS}, impurity\cite{LibischF} and etc. Especially, a recent experiment found that the Fermi velocity (corresponds to hopping energy between nearest neighbor atoms) decreases in the ring area closed to the KQD's boundary\cite{QiaoJB}, due to the strong interaction with the substrate. When the hopping energy is small enough, the KQD will no longer interact with the environment. Therefore, the quasi-bound state will gradually transform into a bound state. However, such process cannot be described by the former method. Thirdly, the shape of KQD is experimentally uncontrollable and a perfect circular shaped QD is difficult to achieve in real experiments. It has been observed that the shape of KQDs can be triangle, trapezoid or rectangle etc\cite{BaiKK,QiaoJB2}. As whispering gallery mode is sensitive to the detailed geometry\cite{StoneAD}, the shape of KQD may greatly affect the quasi-bound state. Moreover, for the non-circular KQD, the Dirac equation is hard to be solved. Therefore, it is useful to find another effective numerical method to overcome these difficulties and analyze the quasi-bound state of KQD quantitatively.

In this paper, we develop a method based on lattice Green\rq{}s Function to calculate the LDOS $\rho$ of KQD in the monolayer graphene. With the help of this method, the LDOS $\rho$ of KQD with arbitrary geometries and non-interaction potential can be accurately obtained without complicate approximation and analytic calculations. From the resonances in LDOS, we can find the quasi-bound states and analyse the confinement effect by the trapping time quantitatively. Three different structures of circular KQD, formed by a step potential, a hyperbolic potential and a combination of hyperbolic potential and hopping, are studied.  The dependence of quasi-bound states on various KQDs parameters are analyzed. For the first and second structures, we find the obtained results not only consist with several experiments \cite{GutierrezC,LeeJ,GhahariF,ZhaoY,BaiKK}, but also approximately agree with the previous theoretical results obtained by solving the Dirac equation, where the intervalley scattering is ignored. However, for the third structure, which has been observed in experiments recently \cite{QiaoJB}, we find the quasi-bound states can turn to real bound states due to the large enhancement of intervalley scattering. The KQDs with different shapes are also studied. The high sensitivity of quasi-bound states to the geometry of KQDs, provides us another way to manipulate quantum states of KQDs.

The rest of the paper is organized as follows. Section II describes the model and the method. Section III describes the specific behavior of LDOS in different KQDs. Section IV gives the conclusion. Finally, some key details of the numerical method are provided in the Appendix.

\section{Model and methods}\label{model}

We consider a KQD distinguished with the surrounding continuous graphene sheet by a different potential, as illustrated in Fig. 1(a). The tight-binding Hamiltonian for the monolayer graphene  is written as:
\begin{equation}\label{H_monolayer}
H=-t\sum_{\langle i,j\rangle }c_i^+c_j+\sum_{\langle i,j\rangle \in \bf{B}}t\rq{}c_i^+c_j+\sum_{i}Uc_i^+c_i \ .
\end{equation}
The first term describes the uniform hopping between the nearest neighbor sites $i$ and $j$. $t$ is the hopping energy.  The second term describes the extra potential modified nearest hopping around the boundary of the KQD. $t\rq{}$ is the modified part of $t$. The last term describes the potential variation due to the KQD. $U$ is the electronic potential.

We adopt the method of the lattice Green's function to calculate the LDOS. Generally, the lattice Green's function applies to the infinite one-dimensional system. To extend it into two-dimensional system, the graphene structure is considered to be periodic in the $x$ direction but be infinite in the $y$ direction. By the method in Appendix \ref{AppendixA}, we obtain the Green\rq{}s function $g^r(E)$ for a square region [e.g. region $L \times W$ in Fig. 1(a)] where a KQD located without considering the potential caused by the KQD.

The potential induced by the KQD contributes a self-energy $\Sigma$, which only contains the nonzero values in a small region. So, one can treat  $\Sigma$ as a perturbation and the final lattice Green\rq{}s function $G^{r}$ at energy $E$  can be calculated by:
\begin{equation}\label{greenfunction}
G^{r}(E)=\{ [g^r(E)]^{-1}-\Sigma \}^{-1}.
\end{equation}
Apart from the Eq. (\ref{greenfunction}), one can also add the self-energy piece by piece to avoid the inversion calculation of a large matrix. The improvement can greatly save the memory of computer.
The details are presented in Appendix \ref{AppendixB}. After obtaining  $G^r$, the LDOS $\rho$ in site $r_i$ can be obtained by:
\begin{equation}
\rho(E,r_i)=-\frac{1}{\pi}{\mathsf {Im}} G^r(E,r_i,r_i),
\end{equation}
where ${\mathsf {Im}} G^r$ represents the imaginary part of  $G^r$.

The LDOS $\rho(E)$ always has some resonances inside the QD. From the resonance features of LDOS, one can find the quasi-bound states and analyse the confinement effect through the trapping time quantitatively. The definition of the trapping time is:
\begin{equation}
\tau=\frac{\hbar}{\delta \epsilon} ,
\label{etau}
\end{equation}
where $\hbar$ is reduced Planck constant and $\delta \epsilon$ is the full-width at half-maximum (FWHM) of the resonance. Based on this definition, the quasi-bound state corresponds to a finite trapping time $\tau$ and the bound state corresponds to a infinite $\tau$. The longer the trapping time $\tau$, the better the confinement effect.

During the numerical calculations, the hopping energy between nearest neighbor sites in a pristine graphene denoted by $t$ is set as the energy unit. Comparing to the previous analytic methods\cite{BardarsonJH, CsertiJ,DowningCA, MatulisA, SchulzQ,SchulzP,WuJS}, the intervalley scattering is included in our numerical calculations. Moreover, our method is applicable for any type of confined potential, thus can be applied to KQD with arbitrary geometry.

\section{numerical results}
With the help of the lattice Green's function method, we obtain the LDOS of  monolayer graphene. In the absence of KQDs, the LDOS are presented in the Appendix \ref{AppendixC}. The results are in agreement with previous studies\cite{NetoAC,NovoselovKS}, which shows strong confirmation of our numerical calculations. In the following, based on such method, we study the properties of KQDs in monolayer graphene.

 \begin{figure}
\center
\includegraphics [width=8.5cm]{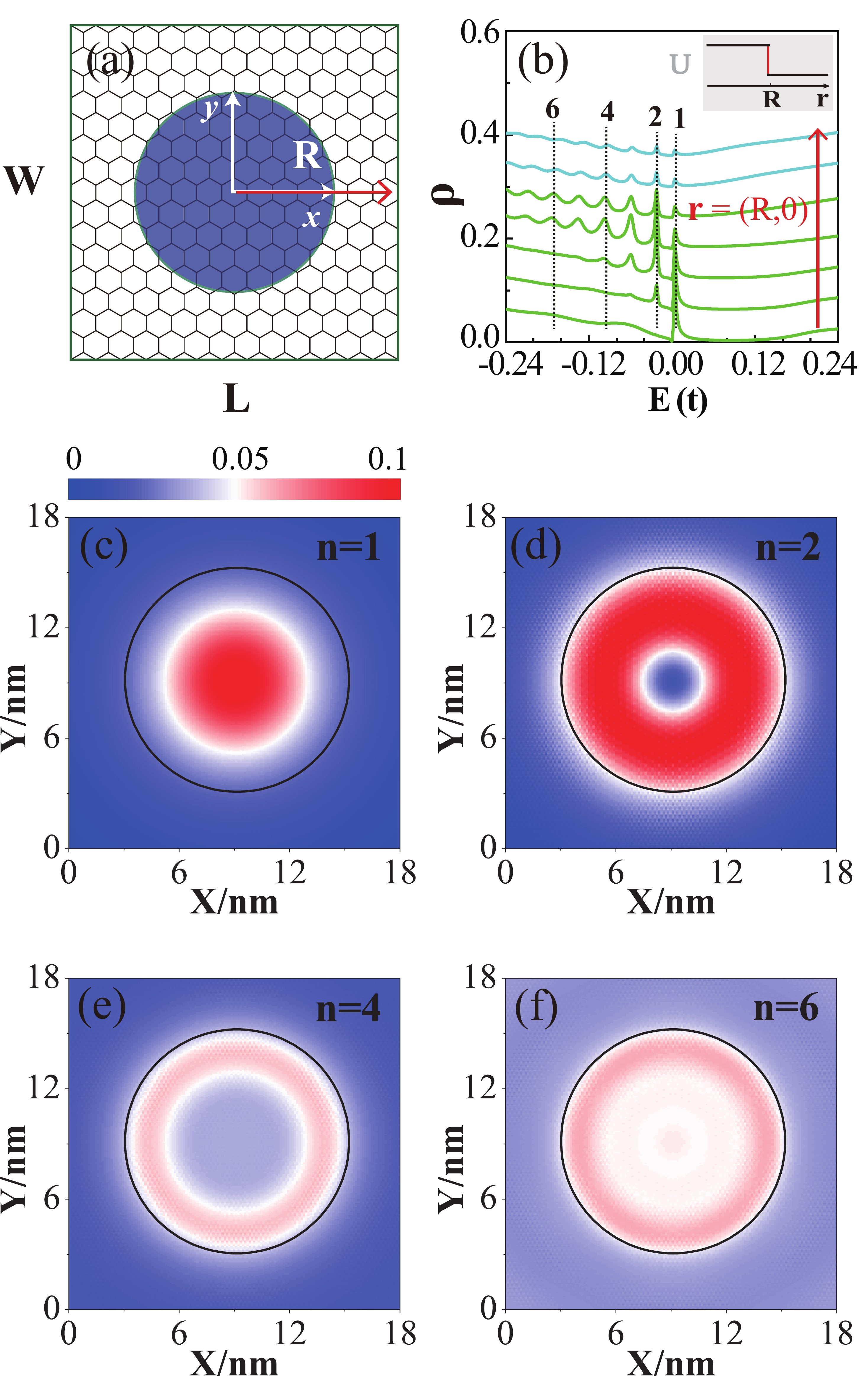}
\caption{(a) Schematic diagram  of the circular graphene KQD with a radius R located in the selected region with length $L=14$ and width $W=8$. The potential has a step change on the boundary, as shown in the inset of (b). (b) The LDOS $\rho$ vs. energy $E$ on different sites along the $x$ direction, which is labelled in (a). The curves are vertically offset for clarity. (c)-(f) Spatial distributions of  LDOS $\rho$ in the energy slices $n=1$, $n=2$, $n=4$ and $n=6$ as labelled in (b). The color bar illustrates the magnitude of $\rho$. The real size of the studied region is $L=75 \ (18 \ nm)$ and $W=43 \ (18 \ nm)$. Other parameters are set as $R=6 \ nm$, and potential $U=V\Theta(R-r)$ with $V=0.1$. \label{QD1}}
\end{figure}

We first study a circular KQD confined by a step potential in monolayer graphene [see Fig. \ref{QD1}(a)]. The studied region is labeled by $L\times W$ with $L=75$ and $W=43$. The lattice constant is set as $a=0.142\ nm$, hence the size of the region is about $18\ nm \times 18\ nm$, which is comparable to the experiments. For better illustration, a rectangular coordinate is set up to label the positions and the origin is located at the center of the selected region. The KQD is also embedded in the center. Here, the potential can be expressed by $U=V\Theta(R-r)$ [see the schematic diagram inset of  Fig. \ref{QD1}(b)], where $\Theta$ is the step function, $V$ is central potential and $R$ is the radius of KQD. Using the method in Sec. II, we obtain LDOS $\rho$ in the selected region. Due to the rotation symmetry of the KQD, the LDOS $\rho$ is isotropic, which is verified by the topographic maps of $\rho$ [see Fig. \ref{QD1}(c)-\ref{QD1}(f)].

\begin{figure}
\center
\includegraphics [width=8.8 cm]{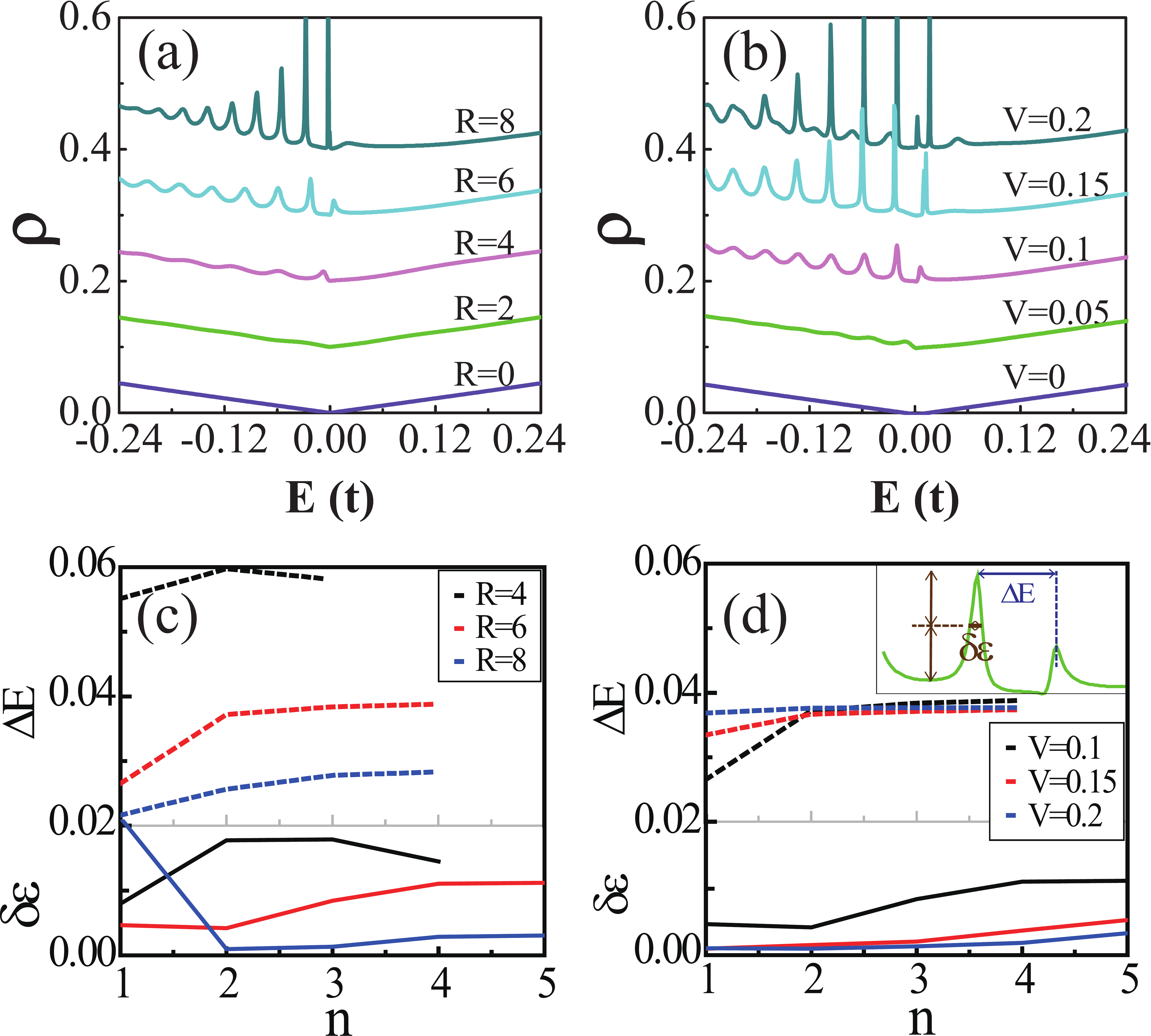}
\caption{(a)(b) The LDOS on boundary $\rho (R)$   vs  energy $E$ under different KQD's radius $R$ (a) and potential energy $V$ (b). (c)(d)  The energy spacing $\Delta E$ (dashed lines) and the FWHM $\delta \epsilon$ (solid lines) of resonances for different energy slices $n$, which correspond to (a) and (b) respectively. The method of measurement ($\Delta E$ with slice $n$ and $\delta \epsilon$ with slice $n+1$) is given by the inset panel (d). The parameters are the same as in Fig. \ref{QD1}. \label{Q1RV}}
\end{figure}

In Fig. \ref{QD1}(b), the  $\rho(E)$ relations on the equally spaced points along red arrow in Fig. \ref{QD1}(a) are plotted. The bottom curve refers to the point close to the center, and the top curve describes the point close to the boundary of the selected region. Here, the green and cyan curves denote the points inside and outside the KQD, respectively. Comparing with the original LDOS plotted in Fig. \ref{C1}(c), we find that $\rho$ is redistributed and many resonance peaks emerges in the energy range $E<0$. Every resonance has a smooth peak, which represents a finite trapping time $\tau$.  Therefore, all the resonances correspond to quasi-bound states. We label these resonance positions with $n=1,2,3,...$ in the energy slice. Interestingly, the resonance peaks on the sites near the boundary of the KQD are the clearest compared to the others. Because of this feature, we choose the edge point [${\bf r}=(R,0)$] for further investigation in Fig. \ref{Q1RV}.

In Fig. \ref{QD1}(c)-\ref{QD1}(f), the LDOS maps at corresponding energy for slices $n=1,~2,~4,~6$ are plotted. For $n=1$, the quasi-bound state is localized in the center of KQD. By increasing $n$, the quasi-bound state gradually moves towards the boundary. For $n=6$, although the confinement of the quasi-bound state becomes weaker due to the increase of energy, the major feature, that the states moving to the boundary, still hold. In the area outside the boundary, $\rho$ also have some resonances, but their amplitudes are small and  decay quickly.  All above features of $\rho$ are observed in the majority of recent experiments of graphene KQDs \cite{GutierrezC,LeeJ,GhahariF,ZhaoY,BaiKK} (e.g. Fig. 2a in ref. \onlinecite{GutierrezC}).

\begin{figure}
\center
\includegraphics [width=8.5 cm]{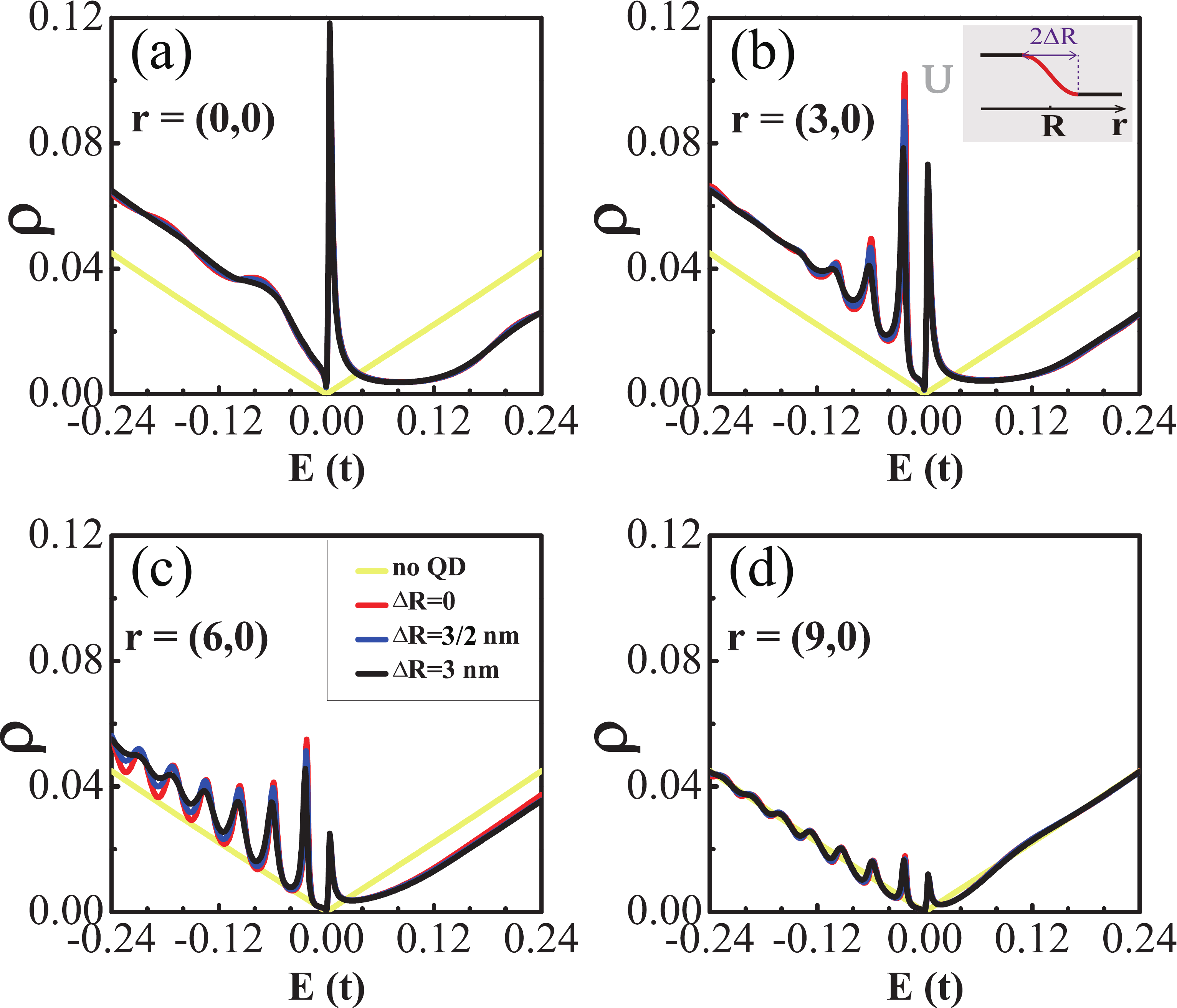}
\caption{The LDOS $\rho$ vs  energy $E$ at the distance (a) ${\bf r}=0\ nm$, (b) ${\bf r}=3\ nm$, (c) ${\bf r}=6\ nm$ and (d) ${\bf r}=9\ nm$ under the smooth potential  $U=\frac{V}{2(1-\tanh \frac{r-R}{S})}$ with variation ranges $\Delta R=0$, $\frac{3}{2}\ nm$, $3\ nm$, which are realized by choosing  $S=0,\ 4a,\ 8a$, respectively.  The inset panel of (b) illustrates the potential profile. For a better comparison,  the case without KQD is also provided (yellow line). Other parameters are $V=0.1$ and $R=6 \ nm$.
\label{QD2}}
\end{figure}

Next, we investigate the effects of central potential $V$ and radius $R$ on the circular KQD. When radius $R$ or central potential $V$ is small, the resonance is indistinct, even on the boundary ${\bf r}=(R,0)$. Only when the radius $R$ or central potential $V$ is large enough, the quasi-bound states can emerge [see  Fig. \ref{Q1RV}(a) and Fig. \ref{Q1RV}(b)]. In order to better characterize the behaviors of these quasi-bound states, we plot their energy spacing $\Delta E(n)=E_{n+1}-E_n$ and FWHM $\delta \epsilon (n)$ in the measurable situations, as shown in Fig. \ref{Q1RV}(c) and Fig. \ref{Q1RV}(d). Here, $E_n$ is the energy of n-th quasi-bound states.   By increasing $R$, both energy spacing $\Delta E$ and FWHM $\delta \epsilon$ decrease significantly. In contrast, when $R$ is fixed, the increasing of  $V$ cannot adjust energy spacing $\Delta E$, although it will decrease FWHM $\delta \epsilon$.  Since the smaller FWHM $\delta \epsilon$ refers to a longer trapping time $\tau$, both the increased radius $R$ and central potential $V$ can enhance the confinement effect.  Moreover, we find energy spacing $\Delta E$ fits the empirical formula, $\Delta E=\frac{3\it{at}}{2\it{R}}$. It is worth to note that the observation $\Delta E=\alpha \it{ \frac{\hbar \upsilon_F}{R}}$$ = \alpha \frac{3\it{at}}{2\it{R}} $ with $\alpha =1$ is reported in recent experiments \cite{LeeJ,GutierrezC}.

In real experiments, the confined potential $U$ is usually not in the manner of step changing\cite{BaiKK2,GhahariF,LeeJ,ZhaoY}. Thus, it is important to investigate the effects of the confined potential type on the quasi-bound states. In Fig. \ref{QD2}, the potential $U$ in a hyperbolic form $U=\frac{V}{2(1-\tanh \frac{r-R}{S})}$ is studied. The smoothness of the potential is characterized by variation range of potential $\Delta R$ [see inset of Fig. 2(b)], which can be simulated by choosing an appropriate $S$. For example, $S=0, ~4a, ~8a$ correspond to $\Delta R=0, ~\frac{3}{2}, ~ 3 ~nm$  with error ratio less than $1\%$. Here, a smaller $\Delta R$ is equivalent to a sharper potential ($\Delta R=0$ correspond to step potential as shown in Fig. \ref{QD1}). Comparing the LDOS $\rho$ on the points ${\bf r}=(0,0)$, $(3,0)$, $(6,0)$, $(9,0)$  for different $\Delta R$.  We find quasi-bound states are nearly insensitive to the variation of $\Delta R$.  For example, the most obvious difference emerges at the edge point ${\bf r}=(6,0)$. Compare $\Delta R =0\ nm$ (red line) and $\Delta R =3\ nm$ (black line) on such point, the resonances of LDOS $\rho$ remain at the same energy and their amplitude differences are very small.

\begin{figure}
\center
\includegraphics [width=8.5 cm]{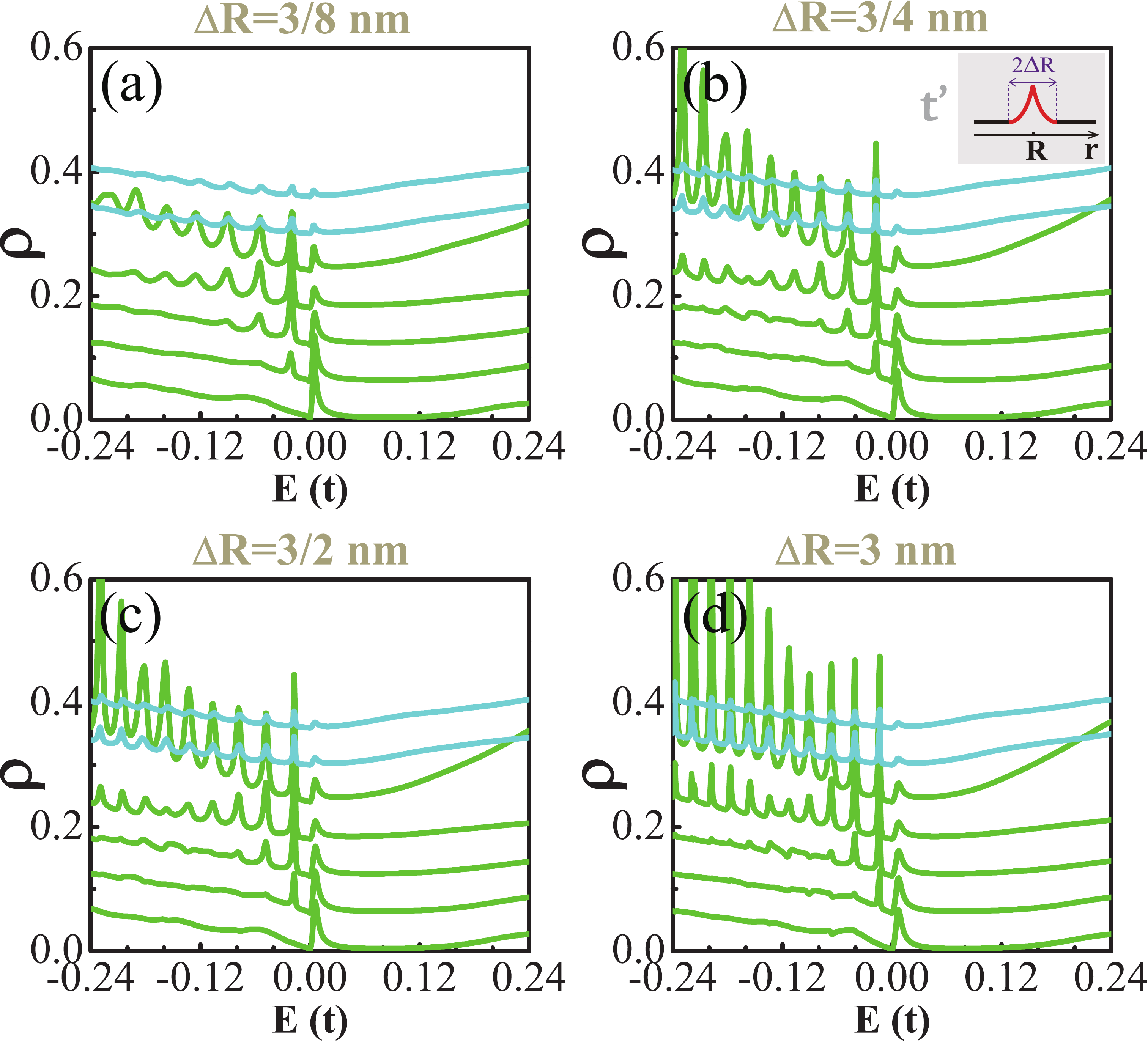}
\caption{The LDOS $\rho$ vs  energy $E$ at the same points of Fig. \ref{QD1}(b) for the combination of smooth potential $U=\frac{V}{2(1-\tanh \frac{r-R}{S})}$ and  hopping   $t'=t_1(1-|\tanh \frac{r-R}{S}|)$  under  variation ranges (a) $\Delta R=\frac{3}{8} \ nm$, (b) $\Delta R=\frac{3}{4} \ nm$, (c) $\Delta R=\frac{3}{2} \ nm$ and (d) $\Delta R=3 \ nm$.  The hopping strength is $t_1=0.5$, and other parameters are the same as Fig. \ref{QD1}.
\label{QD3dR}}
\end{figure}

In most of the previous theoretical studies, the quasi-bound states are obtained by solving continuous Dirac equation under potential $U$, where only one valley is considered. The intervalley scattering, which is inevitable in experiments, is neglected\cite{BardarsonJH, CsertiJ,DowningCA, MatulisA, SchulzQ,SchulzP,WuJS}. However, such different treatment in theory and experiment actually lead to the same results, which has puzzled physicists for a long time\cite{GutierrezC,GhahariF,LeeJ}. In our study, the step potential of $U$ introduces the intervalley scattering while the smoothness of the potential weakens such effect. The results in the above paragraph indicate that the intervalley scattering caused by the potential $U$ is very weak, which verify the validity of previous approximation in several cases.

\begin{figure} \center
\includegraphics [width=8.8 cm] {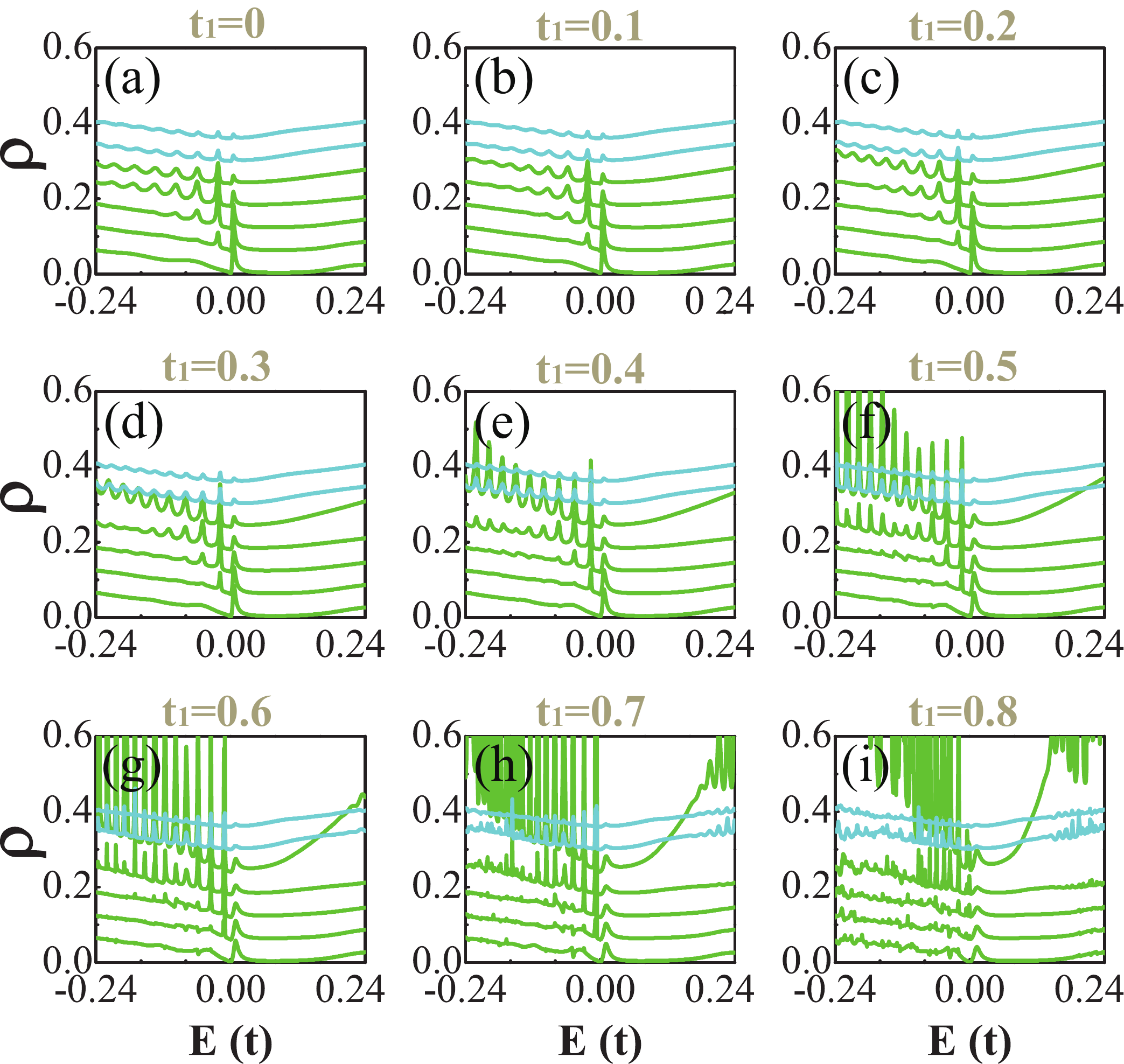}
\caption{The LDOS $\rho$ vs energy $E$ at the same points of Fig. \ref{QD1} (b) for the combination of smooth potential and hopping with  (a) $t_1=0$, (b) $t_1=0.1$, (c) $t_1=0.2$, (d) $t_1=0.3$, (e) $t_1=0.4$, (f) $t_1=0.5$, (g) $t_1=0.6$, (h) $t_1=0.7$ and (i) $t_1=0.8$. Apart from $t_1$, other parameters are fixed as  $\Delta R=3\ nm$, $V=0.1$ and $R=6 \ nm$.
\label{QD3dt}}
\end{figure}

\begin{figure}\center
\includegraphics [width=8.7 cm]{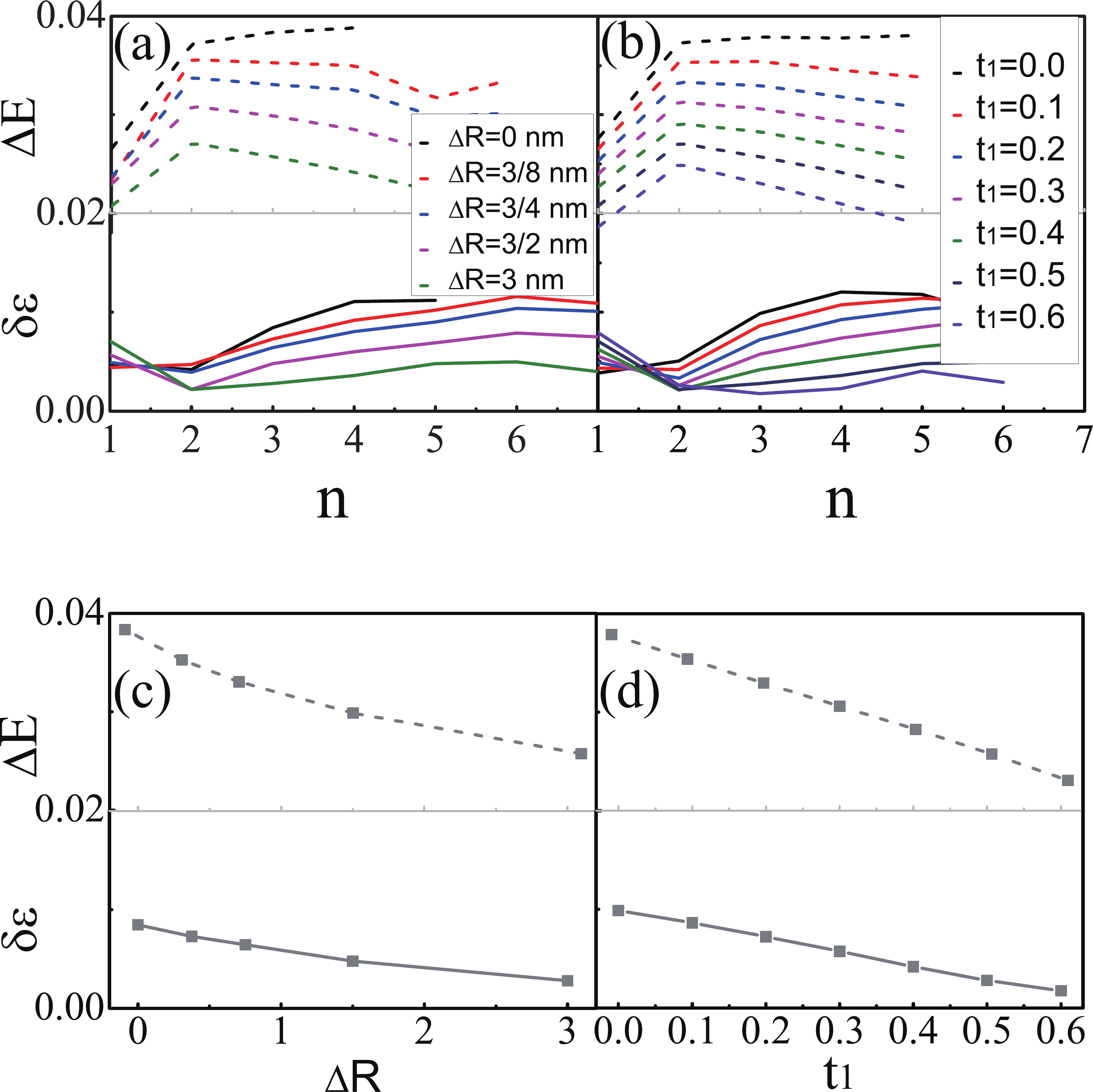}
\caption{Energy level spacing $\Delta E$ (dashed line) and FWHM $\delta \epsilon$ (solid line) of the resonances on the boundary point for different variation range $\Delta R$ ((a) and (c)) and hopping strength $t_1$ ((b) and (d)). The data are collected from Fig. \ref{QD3dR} and Fig. \ref{QD3dt}. $n$ is the sequence of energy slices start from right side. (c) and (d) are the selected data for $n=3$. In (a) and (c), the parameters are $t_1=0.5$, $V=0.1$ and $R=6 \ nm$,. For (b) and (d), the parameters are $\Delta R=3 \ nm$, $V=0.1$ and $R=6 \ nm$.
\label{Q3eE}}
\end{figure}

The KQD structure may not only introduce on site potential $U$, but can also alter the hopping energy between the nearest site in reality \cite{QiaoJB,BaiKK2}. In particular, in one of recent experiments, the decreasing of Fermi velocity in the ring area closed to the KQD boundary is observed\cite{QiaoJB}, which means the hopping energy between the nearest sites decrease from $t$ to $t-t'$. It is worth to note that the decreasing of hopping energy can cause strong intervalley scattering, which is discovered in Dirac systems\cite{zhouJ}.  Nevertheless, the effects of hopping term are not considered in previous graphene KQD studies \cite{BardarsonJH, CsertiJ,DowningCA, MatulisA, SchulzQ,SchulzP,WuJS}. Therefore, we consider the situation that both hopping $t'$ and potential $U$ take hyperbolic form, say,  $t'=t_1(1-|\tanh \frac{r-R}{S}|)$ and $U=\frac{V}{2(1-\tanh \frac{r-R}{S})}$ in the boundary region, respectively, where $t_1$ is the hopping strength. In Fig. \ref{QD3dR}, we focus on the effect of variation range $\Delta R$ on behavior of LDOS $\rho$  under fixed $t_1=0.5$ and $V=0.1$. The approximate variation ranges are $\Delta R=\frac{3}{8}\ nm$, $\frac{3}{4}\ nm$, $\frac{3}{2}\ nm$ and $3\ nm$, corresponding to $S=a$, $2a$, $4a$ and $8a$, respectively. We compare these four plots with Fig. \ref{QD1}(b), which corresponds to the zero variation range ($\Delta R=0$). The observed resonances are still more significant at the edge point. By increasing $\Delta R$, we find the resonances amplitude enhanced and their spacing decreased. In order to show such behaviors more clearly, the energy spacing $\Delta E$ and FWHM $\delta \epsilon$ are plotted in Fig. \ref{Q3eE}(a). For larger $\Delta R$, both $\Delta E$ and $\delta \epsilon$ become smaller.  Since smaller $\delta \epsilon$ means longer trapping time $\tau$, the confinement effect is enhanced by increasing $\Delta R$.  Because it has been verified in Fig. \ref{QD2} that hyperbolic potential $U$ cannot enhance the confinement effect, the hyperbolic hopping $t'$, which brings the intervalley scattering, is the dominant reason for such enhancement \cite{zhouJ}.

The hopping strength $t_1$ defines the coupling strength between KQD and the environment. For example, when $t_1=t$, the KQD becomes an isolated QD.  We investigate the dependence of $t_1$ on LDOS $\rho$ for fixed $\Delta R=3 \ nm$. In Fig. \ref{QD3dt}(a)-\ref{QD3dt}(i), $t_1$ is gradually increased from $0$ to $0.8t$.  The LDOS $\rho$ behaves similarly as the increase of $\Delta R$ (Fig. \ref{QD3dR}). The LDOS $\rho$ on the edge points are significantly changed. When $t_1$ increases, since the energy spacing $\Delta E$ and FWHM $\delta \epsilon$ become smaller [also see  Fig. \ref{Q3eE}(b) and (d)], the resonances of $\rho$ is enhanced. Interestingly, in the condition that the hopping strength $t_1\le 0.6$, one can see the regular resonances in the energy regime $[-0.24t,0]$. However, when $t_1\ge 0.7$, the resonances become chaotic. Moreover, as shown in Fig. \ref{QD3dt}(h) and Fig. \ref{QD3dt}(i), new resonances emerge in the positive energy. Since the LDOS inside the confined region is smaller than that in the environment, it is hard to forming quasi-bound state \cite{BaiKK}. Therefore, the resonances in these two subplots represent real bound states. Figure \ref{QD3dt} demonstrates the evolution from quasi-bound states [Fig. \ref{QD3dt}(a)] to real bound states [Fig. \ref{QD3dt}(g)].

\begin{figure*}\center
\includegraphics [width=15 cm]{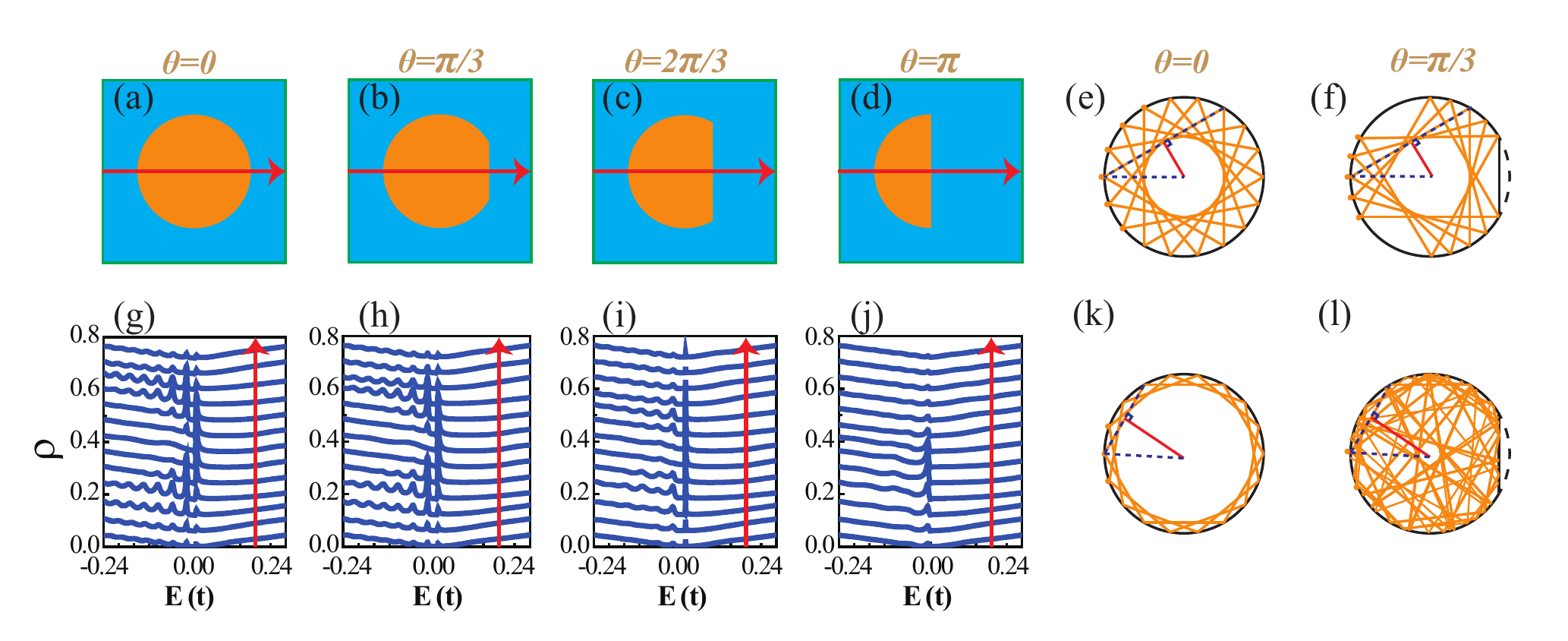}
\caption{(a)-(d) Four typical configurations of KQDs during the evolution from a whole circle (a) to a half circle (d) gradually. (g)-(j) the LDOS $\rho$ vs energy $E$ at the points along direction labelled in (a)-(d) correspondingly. These configurations are marked by the angle $\theta$ of missing parts, i.e., $\theta=0$ [(a) and (g)], $\pi/3$ [(b) and (h)], $2\pi/3$ [(c) and (i)], and $\pi$ [(d) and (j)].  The other parameters are $R=6 \ nm$, $t\rq{}=0$, $U=V\Theta(R-r)$ and $V=0.1$. (e), (f), (k) and (l) illustrate the typical  closed interference loops of reflected carriers inside the KQD for $\theta=0$ [(e) and (k)] and $\pi/3$ [(f) and (l)], respectively.
\label{banyuan}}
\end{figure*}

Figure \ref{Q3eE} shows the energy spacing $\Delta E$ and the FWHM $\delta \epsilon$ of the resonances from the edge point in Fig. \ref{QD3dR} and Fig. \ref{QD3dt}. For most of the resonant slices $n$, both energy spacing $\Delta E$ and FWHM $\delta \epsilon$  have a decrease for increasing variation range $\Delta R$ and hopping strength $t_1$ \cite{footnote}. That is to say, the increasing variation range $\Delta R$ and hopping strength $t_1$ can both enhance the confinement effect of quasi-bound states in the KQD. Furthermore, the typical data of the slice $n=3$ is selected to quantitatively study the enhancement effect in detail [see Fig. \ref{Q3eE}(c) and (d)]. According to Eq. (\ref{etau}), the decreasing of $\delta \epsilon$ from 0.01 to 0.0012 in Fig. \ref{Q3eE}(d) indicates that the trapping time $\tau$ can be adjusted in a large scale.  For energy spacing $\Delta E$, one can fit the data with the empirical formula $\Delta E = \frac{3at^*}{2R^*}$. Here, the parameters $R^*$ and $t^*$ are effective radius and hopping energy, respectively. In contrast, when variation range $\Delta R$ is fixed, we find the relation that $\Delta E \propto (t-t_1)$.  And when the hopping strength $t_1$ is fixed, we find the relationship $\Delta E \propto \frac{1}{0.5R+0.4\Delta R}$. Based on these two relations, we obtain the formula for energy spacing as $\Delta E = \frac{3a}{2} \cdot \frac{(t-t_1)}{ 0.5 R+0.4\Delta R}$. This formula actually tells us that the effective radius decreases in such situation $ (R \ge R^{*}= 0.5 R+0.4\Delta R ) $. Moreover, the effective hopping energy $t^*=(t-t_1)$ equals to the hopping energy between neighbor sites at the boundary, which indicates the quasi-bound states are caused by the mechanisms highly related to the boundary.

Indeed, the above numerical results show a direct evidence that quasi-bound states in KQD originate from a mechanism analogous to whispering gallery mode, where massless electrons are reflected on the boundary and interference with themselves [see Fig. 7(e) and (k)].  As stated in the introduction part, the whispering gallery mode is sensitive to the shape of the structure. In order to further understand such mechanism, it is better to explore the properties of KQD with different shapes. However, due to the difficulty of solving Dirac equation for complicate shapes, such studies are absent before. Here, four shapes of KQD as illustrated in Fig. \ref{banyuan}(a)-\ref{banyuan}(d) are considered and their corresponding LDOS in the transverse directions are plotted in Fig. \ref{banyuan}(g)-\ref{banyuan}(j). When KQD is a perfect circle [see Fig. \ref{banyuan}(a)], we observe many resonances in LDOS. As the KQD evolves from circle shape to semi-circle shape [see Fig. \ref{banyuan}(b)-\ref{banyuan}(d)], these resonances fade away gradually. Moreover, these resonances at high energy fade faster than those with lower energy.  The phenomena can be explained by the whispering gallery mode theory\cite{GhahariF,ZhaoY,ForemanMR,StoneAD}. Specifically, the quasi-bound states at high energy are dominated by whispering gallery mode with high angular momentum. Since the incident angle is large,  the massless electrons will experience multi-reflection processes before forming a closed interference path [see Fig. 7(k)].  In contrast, the quasi-bound states at low energy are dominated by the whispering gallery mode with low angular momentum. After a few reflection processes, the massless electrons can form a closed interference path [see Fig. 7(e)]. When the shape of KQD has a little deformation, e.g. $\theta =\frac{\pi}{3}$, the whispering gallery mode with low angular momentum can still form closed interference path [see fig. 7(f)]. The low energy quasi-bound states still exist. On the contrary, the whispering gallery mode with high angular momentum cannot form a closed interference path [see fig. 7(l)], hance the high energy quasi-bound states fade away. For semi-circle shape, it is hard to form any closed interference path. Thus, there is only one quasi-bound state locating inside the center region. The above study not only reveals the relationship between the whispering gallery modes and the quasi-bound states, but also gives us a way to manipulate the quasi-bound states by controlling the shapes of KQD.

\section{Conclusion}\label{conclusion}
In this paper, we develop a lattice Green's function numerical method, which can be used to calculate the LDOS of KQD with arbitrary geometry and non-interaction potential. We apply this method to monolayer graphene KQD systems, and obtain their LDOS in the presence of various KQD geometries and different types of confined potentials {\textit etc}.  In specific, several calculations are drawn from these numeric results. First, we obtain the behaviors of quasi-bound states in KQD that observed in recent experiments. Second, we show that the reason why circular KQD results in several experiments can be simulated by the previous continuous Dirac equation studies is the weak intervalley scattering. Third, we find the intervalley scattering is greatly enhanced by the confined potential, which reduces the hopping energy. The quasi-bound states can transmit into the bound states. Finally, through the studies of whispering galley mode in different shapes of KQD, we find the quasi-bound states can be manipulated by the geometry of the KQD.

\section{Acknowledgment}\label{acknowledgment}
We thank Lin He, Haiwen Liu for helpful discussion. This work was supported by NSFC under Grants No. 11534001, NSF of Jiangsu Province under Grants No. BK2016007 and NBRPC under Grants 2014CB920901.

\section{Appendix}\label{appendix}

\setcounter{equation}{0}
\setcounter{section}{0}
\renewcommand{\theequation}{A\arabic{equation}}

\subsection{Lattice Green's function for the two-dimensional system} \label{AppendixA}

\begin{figure} 
\center
\includegraphics [width=8 cm]{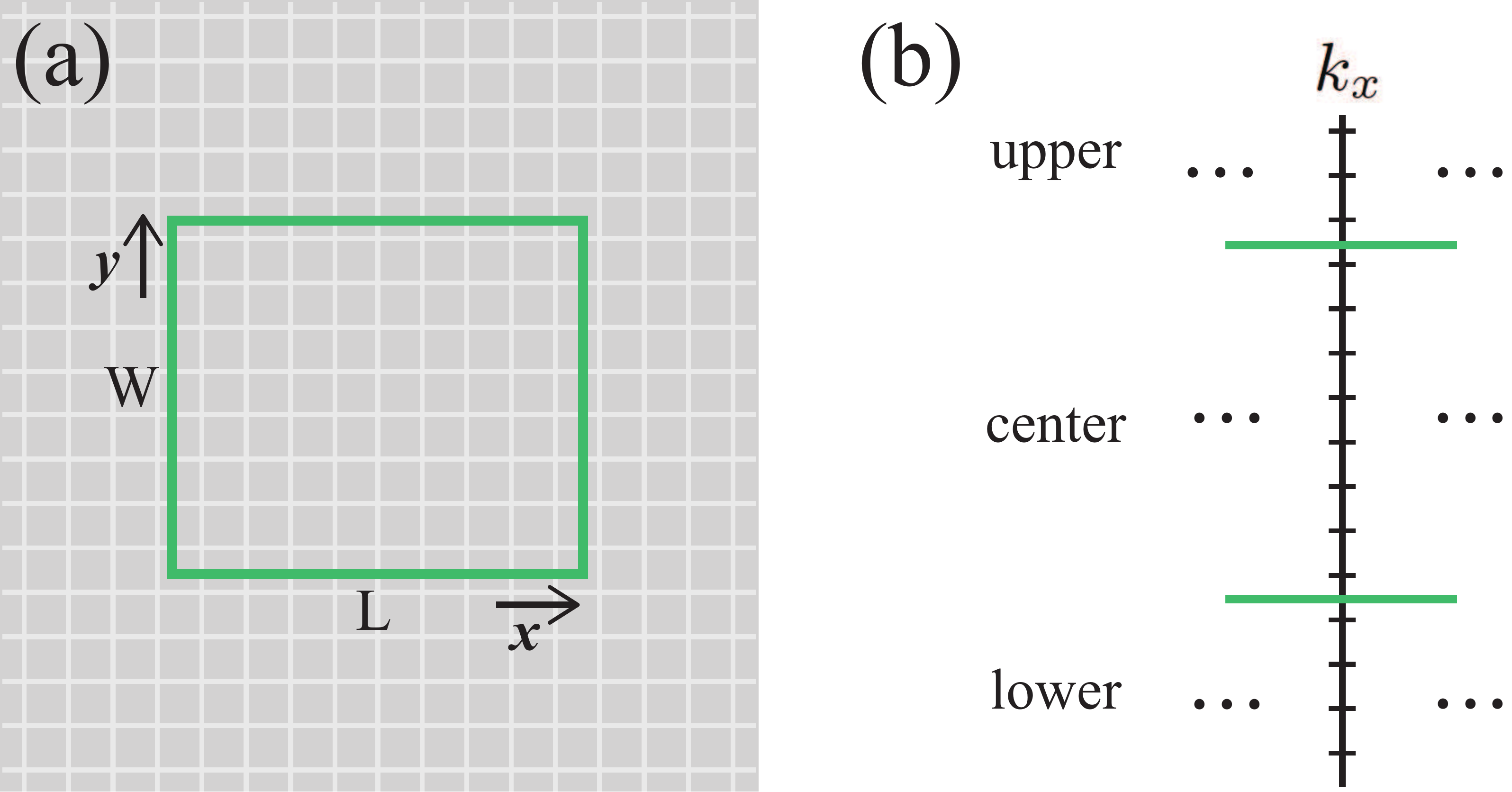}
\caption{(a) Schematic of a selected region $L\times W$ inside an infinite graphene plane. (b) For a centain $k_x$,  the one-dimension chain is separated into three parts: upper and lower semi-infinite leads and central region with length W.
\label{A1}}
\end{figure}

Because KQD is in a small region compared with the whole graphene plane, we only need to calculate the lattice Green's function of this region [see the rectangle with size $L \times W$ in Fig. \ref{A1}(a)]. The whole calculation processes are summarized as follows. Firstly, by applying the Fourier transformation in $x$ direction to Hamiltonian of pristine graphene, one gets a 1D lattice Hamiltonian $\mathcal{H}(k_x)$.  $\mathcal{H}(k_x)$ can be divided into three parts along y direction:  the upper and lower semi-infinite leads and center part with width W, as illustrated in Fig. \ref{A1}(b). Thus, the Green\rq{}s function for a fixed $k_x$ in the center part can be written as:
\begin{equation}
g^r(E, k_x)=[E+i0^{+}-\mathcal{H}_C(k_x)-\Omega_{upper}(k_x) -\Omega_{lower}(k_x)]^{-1},
\end{equation}
where $\mathcal{H}_C(k_x)$ is the Hamiltonian in the center part. $\Omega_{upper} (k_x)$ and $\Omega_{lower}(k_x)$ are the self-energy of upper and lower semi-infinte leads, respectively\cite{LeeDHSanchoMPL}.

Then, the Green's function from position $x_2$ to $x_1$ can be calculated from the integral:
\begin{equation}
g^r (E, x_1,x_2)=\int g^r(E, k_x)e^{ik_x(x_1-x_2)}dk_x,
\end{equation}
Finally, after $x_1, x_2$ takes all the position inside $x_{1,2} \in [1,L]$,  the lattice Green's function $g$ for selected region is obtained by
\begin{equation}
g^{r}(E)=\begin{bmatrix}g^{r}_{11} & \dots & g^r_{1L} \\
\hdotsfor{3} \\
g^{r}_{L1}  & \dots & g^{r}_{LL}
\end{bmatrix}.
\end{equation}

There are two advantages of this method: (i) only one integral process is needed in the whole calculation;  (ii) the accuracy is good enough.  Nevertheless, this method has a shortcoming: the value of $0^+$ cannot be very small, since the interval $dk$ needs to satisfy the condition $dk<0^+/10$ during the integration process.

\setcounter{equation}{0}
\renewcommand{\theequation}{B\arabic{equation}}
\subsection{Effective Green's function algorithm with low computer memory cost}
\label{AppendixB}
In the experiment, the radius of KQD can reach 10nm-20nm \cite{GutierrezC,QiaoJB}. In order to simulate the experimental conditions, the selected region size should be comparable to the size of KQD. That is to say, the size of Hamiltonian matrix should be as large as $60000 \times 60000$, corresponding to the $60000$ atoms in the selected region. In this case, the inversion calculation of Eq. (\ref{greenfunction}) is a challenge due  to the limited memory of computer. Here, we put forward an alternative method to calculate Green's function $G^r$. According to the section II and Appendix A, all lattice Green's function $G^r$, $g^r$ and self-energy $\Sigma$ can be written in a block form $\{  A_{i,j}  \}$  where $i,j \in [1,2, \dots L]$. Obviously, the size of each block $A_{ij}$ is sufficiently small (e.g. for W $ \approx $ 40 \ nm, its size correspond to $380 \times 380$). The method originates from the fact that one can add self energy $\Sigma =\sum_{ij} \Sigma_{ij}$ to Dyson equation $G^r=g^r+ g^r \Sigma G^r$ through block self-energy $\Sigma_{ij}$ piece by piece. The calculation processes are listed as follows. Firstly, for a nonzero block self energy $\Sigma_{mn}$, one can get equation
\begin{equation}
G^r_{nj}=g^r_{nj}+g^r_{nm}\Sigma_{mn}G^r_{nj},\label{B3}
\end{equation}
from Dyson equation. Thus, $G^r_{nj} =[I-g^r_{nm} \Sigma_{mn} ]^{-1} g^r_{nj} $.  Secondly, the block Green's function $G^r_{ij}$ is obtained by:
\begin{equation}
G^r_{ij}=g^r_{ij}+g^r_{im}\Sigma_{mn}G^r_{nj}.
\end{equation}
In Eq. (B1)-(B2), $G$ and $g$ are the Green's function with and without the self-energy $\Sigma_{mn}$.  Thirdly, we replace $g^r$ by $G^r$ and consider another
nonzero $\Sigma_{mn}$. When all nonzero $\Sigma_{mn}$ are counted, the finally Green's function $G^r$, which is equivalent to that in Eq. (2), is obtained.

During the calculation processes, the memory cost is quite low.  Note that the number of nonzero element $\Sigma_{ij}$ is usually small, so that such algorithm can increase computation efficiency.

\begin{figure}
\center
\includegraphics [width=8.5 cm]{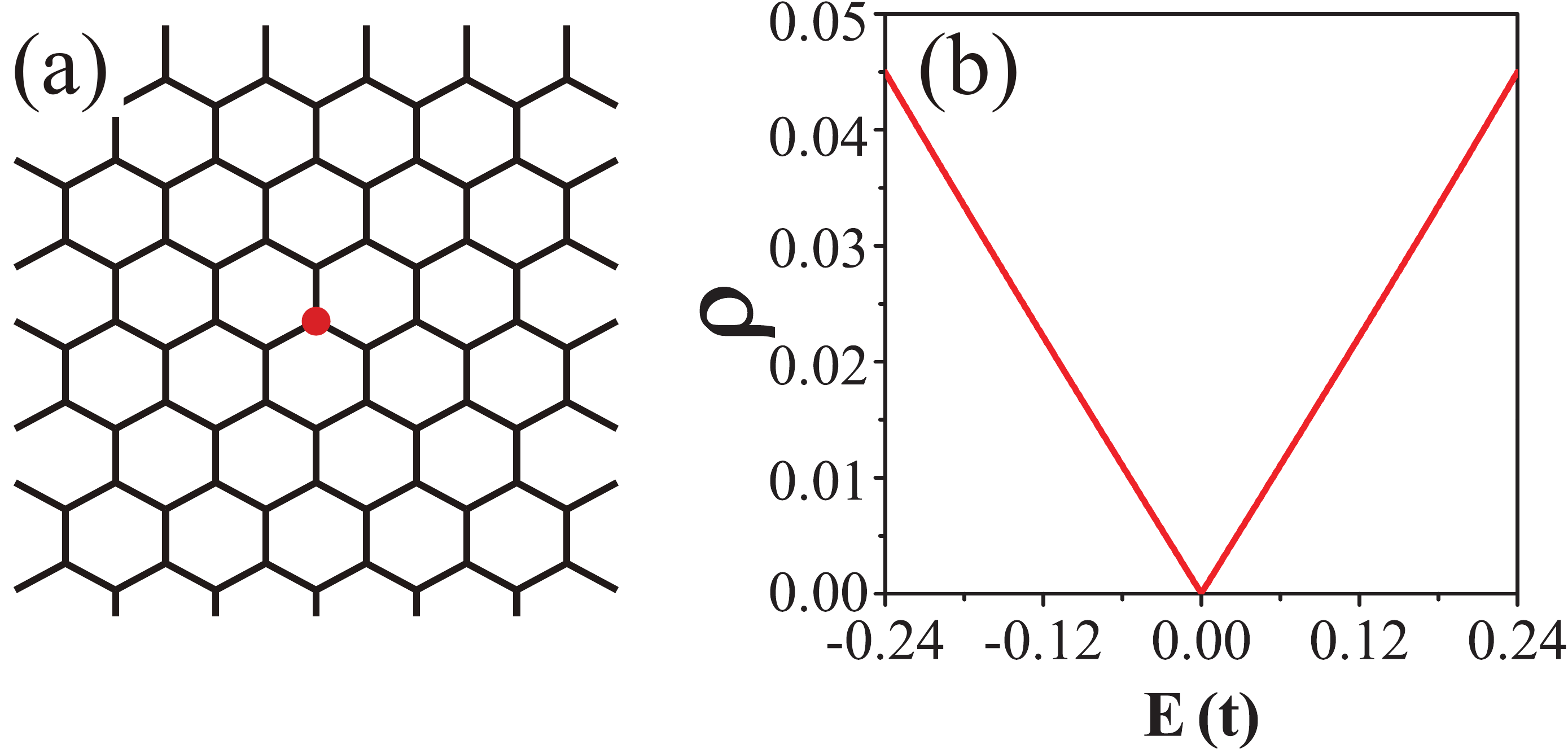}
\caption{(a) is the schematic diagram of monolayer graphene plane. (b) is the LDOS $\rho$ varied with energy $E$ on the position labeled in the panel (a).
\label{C1}}
\end{figure}

\setcounter{equation}{0}
\renewcommand{\theequation}{C\arabic{equation}}
\subsection{LDOS in monolayer graphene without KQD}
\label{AppendixC}

In order to test the accuracy of the method, we try to obtain the LDOS in monolayer graphene without the KQD by applying this method. The sites in monolayer graphene are equivalent and  their LDOS behaviors are the same as the red line in Fig. \ref{C1}(c). Here, the LDOS reproduces the same behaviors in ref. \onlinecite{NetoAC, NovoselovKS}. Moreover, the LDOS satisfy $\rho(E)=\rho(-E)$, showing the particle-hole symmetry of both systems. $\rho$ increases linearly with the energy $|E|$, which is consistent with the linear dispersion near the Dirac point.


\begin{thebibliography}{99}
\bibitem{footnote}
For $n=1$, we find that FWHM $\delta \epsilon$ monotonous increase with $\Delta R$ and $t_1$. The quasi-bound state of $n=1$ is mainly localized in the center of KQD, while our data are collected from the LDOS on edge point. The departure from the mainly localized position may cause the inaccuracy of the $\delta \epsilon$ and leads to such anomalous phenomenon.
\end{thebibliography}
\end{document}